\documentclass[aps,prl,twocolumn,groupedaddress,showpacs]{revtex4}
\usepackage{graphicx}
\usepackage{times}

\begin{document}
\title{Suppression of friction by mechanical vibrations}

\author{Rosario Capozza$^{1}$
Andrea Vanossi$^{2,1}$,Alessandro Vezzani$^{1,3}$ and Stefano Zapperi$^{1,4}$}
\affiliation{$^{1}$CNR-INFM, S3, Dipartimento di Fisica,
Universit\`a di Modena e Reggio Emilia, Via G. Campi 213A, 41000 Modena, Italy\\
$^2$ International School for Advanced Studies (SISSA)
and CNR-INFM Democritos National Simulation Center, Via Beirut 2-4, I-34014 Trieste, Italy\\
$^{3}$Dipartimento di Fisica, Universit\`a di Parma, Parma, Italy\\
$^{4}$ISI Foundation, Viale S. Severo 65, 10133 Torino, Italy}

\date{\today}

\begin{abstract} Mechanical vibrations are known to affect frictional sliding
and the associated stick-slip patterns causing sometimes a drastic reduction of the friction force. This issue is relevant for applications in nanotribology and to understand earthquake triggering by small dynamic perturbations .
 We study the dynamics of repulsive particles confined between a horizontally driven top plate and a vertically
 oscillating bottom plate. Our numerical results show a suppression of the high dissipative stick-slip regime in a
 well defined range of frequencies that depends on the vibrating amplitude, the normal applied load, the system
 inertia and the damping constant. We propose a theoretical explanation of the numerical results and derive a phase
 diagram indicating the region of parameter space where friction is suppressed. Our results allow to define better
 strategies for the mechanical control of friction.
\end{abstract}

\pacs{81.40.Pq, 46.55.+d, 68.35.Af, 68.08.-p}

\maketitle

Natural or artificially induced manipulations by small mechanical
vibrations, when applied at suitable frequency and amplitude ranges,
may help in driving a contacting sliding interface
out of its potential energy minima, thus increasing considerably
surface mobility and diffusion, and reducing friction. This has been shown experimentally for sliding of nanoscale contacts through, e.g., the atomic force microscope \cite{socoliuc06,jeon06,su03}, and in computer simulations via extended molecular dynamics \cite{gao98} and simple modeling approaches \cite{rozman98,zaloj99,tshiprut05}. On a larger scale, it has been observed that in sheared granular media experiments the stick slip behavior is significantly perturbed by tiny transverse vibrations \cite{johnson05,johnson08}. Since geological faults are often filled with a granular gouge, these results might be relevant to understand earthquake triggering by low amplitude seismic waves \cite{stacey05}. Despite these promising numerical and experimental contributions, a quantitative theory accounting for the friction dependence on vibrations is still lacking.

\begin{figure}
\includegraphics[width=1.\linewidth]{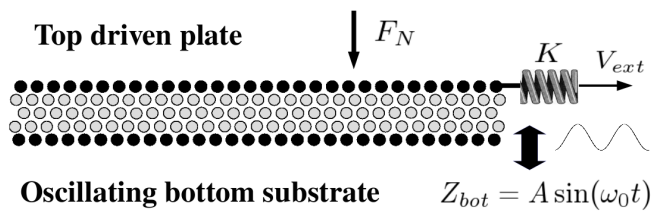}
\caption{Sketch of the system, with the rigid top and bottom plates
indicated in black and the confined particles in red. The top is
dragged through a spring of elastic constant $K$ moving at
constant velocity $V_{ext}$, while the bottom vibrates vertically with
frequency $\omega_0$ and amplitude $A$}
\label{system}
\end{figure}

In this letter, we study the frictional properties of a two dimensional set of repulsive particles 
confined between two rigid plates (see Fig.\ref{system}) \cite{braun01,braun04}. The top plate
is attached to a spring that is pulled at constant velocity, while the bottom plate is vibrated vertically. 
Without vibration, the top plate would slide exhibiting a characteristic stick-slip behavior. Vibrations
induce a drastic reduction of the friction coefficient and a suppression of the stick-slip behavior, but only
in a well defined frequency range. We propose a theoretical argument to explain this behavior and construct
a phase diagram indicating the parameter region for which friction is suppressed. The theoretical results are in
excellent quantitative agreement with numerical simulations. Finally, we investigate the frictional response
of the system to a short vibration pulse and find large slip events when the vibration frequency lies in the appropriate regime. These results could be relevant to understand triggering of frictional slip by small perturbations.

Our two dimensional system consists of two identical top and bottom
substrates, composed of $n_t=n_b$ particles with coordinates ${\bf r}_i^t$ and ${\bf r}_i^b$ respectively and constant lattice separation $a_s$=1. We confine $n_{p}$ particles with coordinates ${\bf r}_i^p$ between the top and bottom plates. The mass $m$ of all
the particles is the same, so that the total mass of the confined layers is $M_{p}=m \cdot n_{p}$ while the top and bottom plate have a mass $M_{bot}=M_{top}=m \cdot n_{t}$. Due to the vertical confinement, periodic boundary conditions are applied only along the $x$ direction. The particles interact via a pairwise repulsive potential $U(r)=U_0\left[\left (\frac{r_0}{r} \right)^{12}-2 \left(\frac{r_0}{r} \right)^{6}\right]$ for $r<r_0$ and $U(r)=0$ otherwise. The parameters of the potential are the same for all the particles, wether in the top and bottom plate or in between them. We adjust the number of particles $n_p$, imposing that $n_p=N_l \cdot n_t=N_l \cdot n_b$ with $N_l$ an integer number. In this way the systems forms $N_l$ perfectly ordered layers of confined particles. We also consider the case of a few missing particles, finding exactly the same results as long as the system remains ordered. The top plate of mass is subject to a normal force
$F_N$ and is pulled through a spring, of elastic constant $K$, 
which moves along the horizontal $x$ direction with constant velocity $V_{ext}$

Indicating with ${\bf R}_{top}=(X_{top},Z_{top})$ and
${\bf R}_{bot}=(X_{bot},Z_{bot})$ the center of mass coordinates of
the top and bottom plate respectively, where
${\bf R}_{top}=\frac{1}{n_{t}}\sum\limits_{i= 1}^{n_{t}}{\bf r}_i^{t}$, and ${\bf R}_{bot}=\frac{1}{n_{b}}\sum\limits_{i=1}^{n_{b}}{\bf r}_i^b$,
the particles satisfy the equations of motion
\begin{eqnarray} \nonumber
m \ddot{{\bf{r}}}_i^p &+& \sum_{i \ne j}^{N}\frac{d}{d{\bf{r}}_i}
U(|{\bf{r}}_{i}^p-{\bf{r}}_{j}|)+m\eta(\dot{{\bf{r}}}_i^p-\dot{{\bf{R}}}_{top})+
\\ \label{lubeq}
&+&m\eta(\dot{{\bf{r}}}_i^p-\dot{{\bf{R}}}_{bot})+{\bf{f}}^{ran} = 0
\\ \nonumber
M_{top} \ddot{X}_{\rm top}&+& \sum\limits_{i= 1}^{n_{t}}\sum\limits_{j=
1}^{n_{p}} \frac{d}{dx_{i}^t}
U(|{\bf{r}}_{j}^{p}-{\bf{r}}_{i}^{t}|) + K(X_{\rm top}-V_{\rm ext}t)+
\\ \label{xtop}
&+& \sum\limits_{i= 1}^{n_{p}}m \eta ({\dot X}_{\rm top}-{\dot x}_{i}^p)
+ f_x^{ran}=0
\\ \nonumber
M_{top} \ddot{Z}_{\rm top} &+& \sum\limits_{i= 1}^{n_{t}}\sum\limits_{j=
1}^{n_{p}}\frac{d}{dz_{i}^{t}}U(|{\bf{r}}_{j}^{p}-{\bf{r}}_{i}^{t}|) +
F_N+
\\ \label{ztop}
&+&\sum\limits_{i= 1}^{n_{p}}m \eta ({\dot Z}_{\rm top}-{\dot z}_{i}^p)
+ f_z^{ran}=0
\end{eqnarray}
where $N_p+n_t+n_b$ and $\eta$ is the damping coefficient that
accounts for a viscous dissipation. The temperature is controlled by a Langevin thermostat according to the relation $\langle {\bf f}(t)^{ran}{\bf f}(t')^{ran}\rangle = 4 m\eta k_B T\delta(t-t')$. In the present simulations, we consider very low temperatures $k_BT=10^{-2}U_0$. 

To study the influence of mechanical vibrations on the system, the bottom plate is vibrated vertically $Z_{bot}=Z_0+A\sin(\omega_0t)$ where $Z_0$ is a reference coordinate, $A$ is the amplitude and $\omega_0$ the frequency, while its horizontal component $X_{bot}$ is held fixed.  In all the simulations, we compute the instantaneous friction force $F_L$ by measuring the spring elongation of the driving apparatus, $(X_{top}(t)-V_{ext}t)$, so that $F_{L} = K\;(X_{top}(t)-V_{ext}t)$. The friction coefficient is defined as $\mu\equiv F_L/F_N$ and its average $\langle\mu\rangle$ , obtained integrating its value over a sufficiently long time interval in the steady state.

\begin{figure}
\includegraphics[width=.9\linewidth]{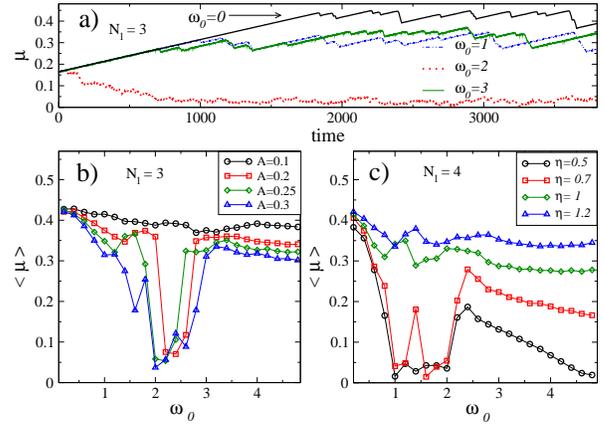}
\caption{a) Friction coefficient $\mu$ vs time obtained for
three values of the bottom oscillating frequency $\omega_0$ (three
lubricant layers). b) Time-averaged value of $\mu$ 
as a function of $\omega_0$ for four values of the oscillation amplitude $A$ (ranging from 3\% to 9\% of the  $N_l=3$ lubricant thickness). c) Time-averaged value of $\mu$ vs $\omega_0$ for $N_l=4$ and four values of the damping coefficient $\eta$.}
\label{fs_om}
\end{figure}

In Fig. \ref{fs_om}a, we show the behavior of the friction coefficient $\mu$ as a function of time in the case of three confined particle layers for three distinct values of the frequency $\omega_0$. For $\omega_0=1$ and $\omega_0=3$, we observe a characteristic stick slip behavior, with loading phases where the system is stuck, followed by rapid slip events in which the force accumulated by the spring is relaxed. A very similar pattern also takes place in absence of vibrations at low external driving rates, with the system alternately sticking and slipping forward (not shown). At $\omega_0=2$, we observe a strong reduction of the friction coefficient and a drastic suppression of the sawtooth stick-slip behavior. To make this observation more quantitative, we report in Fig. \ref{fs_om}b and \ref{fs_om}c the systematic variations of the average friction coefficient $\langle\mu\rangle$ with the vibration frequency $\omega_0$. The left panel shows results obtained for various oscillation amplitudes $A$, ranging from 3\% to 9\% of the film thickness, while the right panel displays results for different values of the damping coefficient $\eta$. We see that the suppression of friction appears in a well defined range of frequencies [$\omega_1,\omega_2$], which depends on $A$ and $\eta$.

\begin{figure}
\includegraphics[width=.9\linewidth]{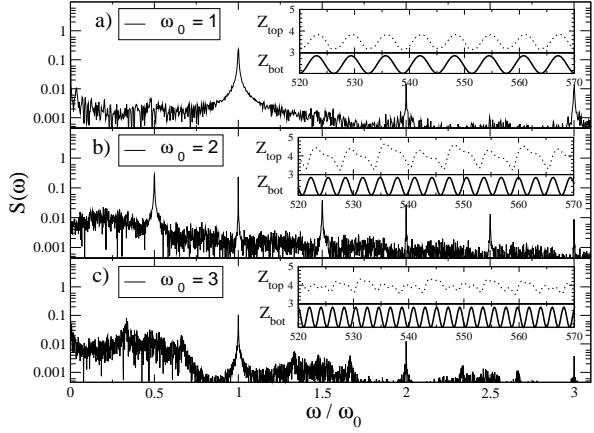}
\caption{Power spectrum of the vertical component $Z_{top}$ of the plate for different values
of the oscillating frequency $\omega_0$ and $N_{l}=3$, $A=0.3$, $\eta=1$. At low frequency ($\omega_0$=1) the system follows the oscillating bottom substrate. At higher frequencies $\omega_0=2$ the system oscillates also at one-half frequency of $\omega_0$. At $\omega_0=3$ a low frequency noise appears.  In the insets we report the time evolution of $Z_{top}$ (dotted line) and $Z_{bot}$ (thick line).}
\label{four}
\end{figure}

To understand the origin of this phenomenon, we compute the power
spectrum of the vertical position of the top substrate $Z_{top}$ for
the three values of $\omega_0$ corresponding to Fig.\ref{fs_om}a.
At low frequencies (inset in Fig.\ref{four}a), the top plate and the confined particles vibrate in phase with the oscillations of the bottom plate. Hence, the spectrum displays a peak at frequency $\omega_0$ and its integer multiples and vibrations have no visible effect on sliding friction. The situation is drastically different at $\omega_0=2$ (Fig.\ref{four}b). In this case, the top plate and the confined particles can not follow the bottom plate. The vertical position of the top plate increases, presenting high amplitude oscillations which diminish the contact time between the confined particles and the bottom plate, reducing considerably the friction force. The corresponding Fourier spectrum shows additional peaks at integer multiples of $\omega_0/2$. We find that this is a general feature of the spectrum in the interval of friction suppression [$\omega_1,\omega_2$]. Additional peaks may appear at frequencies $\omega=\omega_0/n_0$, where $n_0$ is an integer, implying that the top plate oscillates with a period that is an integer multiple of the driving period on the bottom plate. Further increases of the oscillation frequency induce a reduction of the amplitude of the top plate oscillations around the equilibrium position (see Fig.\ref{four}c), so that the low frequency noise suppresses the resonant peaks. As a result, the friction force increases again.

The mechanism illustrated by the spectral analysis suggests an argument to derive analytically the values of $\omega_{1}$ and $\omega_{2}$, defining the interval of friction suppression.
At low frequencies, the particles follow coherently the oscillation of the bottom plate owing to the downward action of the normal load $F_N$. The reduction of friction occurs when the particles are able to detach from the bottom plate due to the action of the inertial force induced by the external vibration $F_{in}=M\ddot{Z}_{bot}$, where $M=M_{p}+M_{top}$. The frequency $\omega_1$ corresponds to the condition for which the inertial force overcomes the combined action of the normal load $F_{N}$ and the damping force $F_{damp}=M_{p} \eta \dot{Z}_{bot}$
\begin{equation}\label{condb1}
F_{in}(\omega_1) \simeq F_{N}(\omega_1) + F_{damp}(\omega_1).
\end{equation}
Estimating the inertial force as $F_{in}\simeq MA\omega_0^2$ and
the damping force as $F_{damp}\simeq M_{p}\eta A \omega_0$,
we obtain an implicit expression for the starting frequency of friction suppression $\omega_1$
\begin{equation}
M A \omega_1^2 = F_{N} + M_{p} \eta A \omega_1 \label{esomin}.
\end{equation}
It is convenient to work with dimensionless quantitities, defining
\begin{equation}\label{adim}
\tilde{f}\equiv\frac{F_{N}}{MA\eta^2}\;\;\;\;
\tilde{m}\equiv\frac{M_{p}}{M}\;\;\;\;
\tilde{\omega}\equiv\frac{\omega}{\eta}.
\end{equation}
Using these rescaled variables, Eq.~\ref{esomin} yields:
\begin{equation}\label{tesomin}
\tilde{\omega}_1=\frac{1}{2}\left(\tilde{m}+\sqrt{\tilde{m}^2+4\tilde{f}}\right)
\end{equation}

To estimate the recovery frequency $\omega_{2}$, we determine the conditions for the presence of low frequency vibrations. 
Due to the external oscillations, the confined layers detach from the bottom substrate during a characteristic time 
$\Delta t$ that can be estimated as
\begin{equation}\label{risetime}
\Delta  t \simeq \dot{Z}_{bot}M/F_{N}\simeq A\omega_0M/F_{N}
\end{equation}
To observe low frequency vibrations, which will cause the recovery of the
friction force, the period of the external oscillation should be smaller than the rise time associated with the internal vibrations
of the particles (i.e. $\frac{2\pi}{\omega_0}<\Delta t$). This condition corresponds to the maximum of the momentum transfer from the
vibrating plate to the confined particles. Using again the dimensionless variables defined in Eq.~(\ref{adim}), we estimate
\begin{equation}\label{tesomend}
\tilde{\omega}_2=\sqrt{2\pi \tilde{f}}.
\end{equation}

\begin{figure}
\includegraphics[width=1.\linewidth]{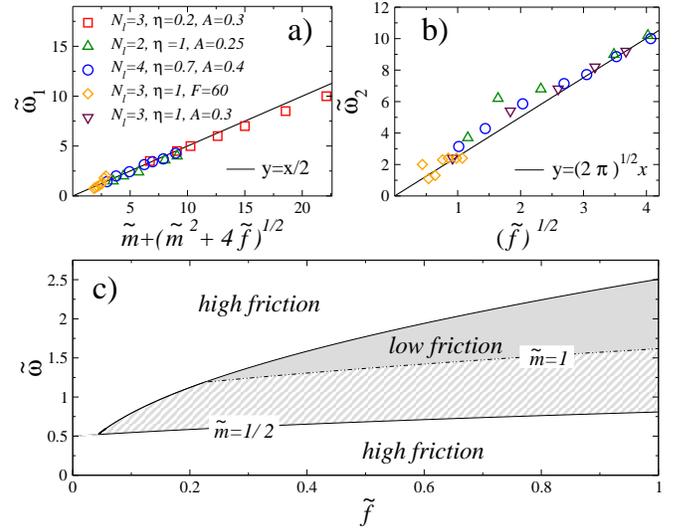}
\caption{ a) Comparison between the numerical results (symbols) and
the theory (solid line). $\tilde{\omega}_1$ as a function of
the dimensionless variables $\tilde{f}$ and $\tilde{m}$.
b) $\tilde{\omega}_2$ as a function of $\tilde{f}$. The behaviors
corresponding to different symbols are obtained keeping
three parameters fixed and varying the fourth.
c) Phase diagram indicating the region of friction suppression.
Top curve: $\tilde{\omega}_2$ vs $\tilde{f}$. Grey-striped region:
$\tilde{\omega}_1$ vs $\tilde{f}$ for
$\frac{1}{2}<\tilde{m}<1$. Friction is strongly suppressed for
$\tilde{\omega}_1<\tilde{\omega}<\tilde{\omega}_2$ (grey region).}
\label{phase_om}
\end{figure}

The theoretical predictions for $\omega_1$ and $\omega_2$ are in excellent agreement with the numerical simulations, as shown in Fig. \ref{phase_om}a and \ref{phase_om}b. The numerical values are obtained varying the number of layers $N_l$, the vibration amplitude $A$, the damping coefficient $\eta$ and the normal load $F_N$. Notice that the theory has no adjustable parameters.

From the analytical relations (\ref{tesomin}) and (\ref{tesomend}), we can draw a phase diagram indicating, in the space of dimensionless
variables (\ref{adim}), the region where friction is suppressed
(Fig.\ref{phase_om}c). The region of friction reduction is enclosed
between $\tilde{\omega}_1$ and $\tilde{\omega}_2$ and shrinks as we
reduce $\tilde f$, until it finally disappears. Notice that $\tilde{\omega}_1$ depends on the value of reduced mass which lies between $\tilde{m}=1/2$, for a single confined layer, and $\tilde{m}=1$ for an infinitely wide system. The possible values for the ${\omega}_1$ are indicated by the grey-striped region in Fig.\ref{phase_om}c.

\begin{figure}
\includegraphics[width=.9\linewidth]{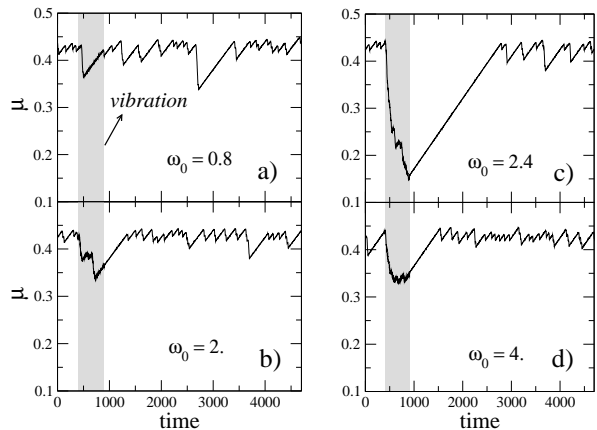}
\caption{The effect of a short time vibration pulse on the stick-slip behavior
for different frequencies $\omega_0$, $N_l=3$, $A=0.2$.
The shaded light grey regions indicate the duration of
vibration pulse. The largest effect is visibile when the frequency lies in the
range of $\omega_1<\omega_0<\omega_2$.}
\label{slip}
\end{figure}

The phase diagram displayed in Fig.\ref{phase_om}c indicates the frictional behavior of the system under a steady vertical oscillation. There are cases, however, where the external vibration acts only for a small amount of time, such in the case of earthquakes or avalanches triggered by seismic waves. To address this issue, we analyze changes in the stick-slip pattern for small vertical vibration of finite duration $T_v$ and frequency $\omega$, with $T_v \gg 1/\omega_0$. To avoid discontinuities, we switch the perturbation on the bottom plate smoothly: $Z_{bot}=Z_0+f(t,T_v)A\sin(\omega_0t)$, where $f(t,T_v)=(\tanh(t/\tau)-\tanh((t-T_v)/\tau))/2$, with
$\tau \ll T_v$. If we chose $T_v$ to be of the same order of magnitude as the stick time, we observe that the perturbation typically leads to small changes in the slip patterns.  When $\omega_1 <\omega_0 <\omega_2$, however, the systems exhibit a large slip event (see Fig.~\ref{slip}). After the perturbation is removed the system recovers the original stick-slip behavior, without any long-range memory effects. Our results suggest that catastrophic events are more likely to be triggered when the perturbation lies in a definite frequency interval.

In conclusions, we have clarified the role of vibrations in the frictional sliding of a confined system. The general mechanism for friction suppression that we have uncovered is based on the reduction of the effective interface contacts produced by vibrations.
Since the results depend only on the relation between inertial and dissipative forces, we expect them to be valid for a wide class of sliding systems, including granular media and nanoscale interfaces. Further work in this direction could be useful to optimize friction control in technological nanodevices and to design better strategies to forecast the triggering of instabilities in materials and geosystems.

{\bf Acknowledgments -} This work is supported by the European Commissions NEST Pathfinder programme TRIGS under contract NEST-2005-PATH-COM-043386, and partially by CNR, as part of the European Science Foundation EUROCORES Programme FANAS.

\vspace{1cm}


\begin{thebibliography}{12}
\expandafter\ifx\csname natexlab\endcsname\relax\def\natexlab#1{#1}\fi
\expandafter\ifx\csname bibnamefont\endcsname\relax
  \def\bibnamefont#1{#1}\fi
\expandafter\ifx\csname bibfnamefont\endcsname\relax
  \def\bibfnamefont#1{#1}\fi
\expandafter\ifx\csname citenamefont\endcsname\relax
  \def\citenamefont#1{#1}\fi
\expandafter\ifx\csname url\endcsname\relax
  \def\url#1{\texttt{#1}}\fi
\expandafter\ifx\csname urlprefix\endcsname\relax\def\urlprefix{URL }\fi
\providecommand{\bibinfo}[2]{#2}
\providecommand{\eprint}[2][]{\url{#2}}

\bibitem[{\citenamefont{Socoliuc et~al.}(2006)\citenamefont{Socoliuc, Gnecco,
  Maier, Pfeiffer, Baratoff, Bennewitz, and Meyer}}]{socoliuc06}
\bibinfo{author}{\bibfnamefont{A.}~\bibnamefont{Socoliuc}},
  \bibinfo{author}{\bibfnamefont{E.}~\bibnamefont{Gnecco}},
  \bibinfo{author}{\bibfnamefont{S.}~\bibnamefont{Maier}},
  \bibinfo{author}{\bibfnamefont{O.}~\bibnamefont{Pfeiffer}},
  \bibinfo{author}{\bibfnamefont{A.}~\bibnamefont{Baratoff}},
  \bibinfo{author}{\bibfnamefont{R.}~\bibnamefont{Bennewitz}},
  \bibnamefont{and} \bibinfo{author}{\bibfnamefont{E.}~\bibnamefont{Meyer}},
  \bibinfo{journal}{Science} \textbf{\bibinfo{volume}{313}},
  \bibinfo{pages}{207} (\bibinfo{year}{2006}).

\bibitem[{\citenamefont{Jeon et~al.}(2006)\citenamefont{Jeon, Thundat, and
  Braiman}}]{jeon06}
\bibinfo{author}{\bibfnamefont{S.}~\bibnamefont{Jeon}},
  \bibinfo{author}{\bibfnamefont{T.}~\bibnamefont{Thundat}}, \bibnamefont{and}
  \bibinfo{author}{\bibfnamefont{Y.}~\bibnamefont{Braiman}},
  \bibinfo{journal}{Appl. Phys. Lett.} \textbf{\bibinfo{volume}{88}},
  \bibinfo{pages}{214102} (\bibinfo{year}{2006}).

\bibitem[{\citenamefont{Su et~al.}(2003)\citenamefont{Su, Xu, Kurita, Kato, and
  Adachi}}]{su03}
\bibinfo{author}{\bibfnamefont{L.}~\bibnamefont{Su}},
  \bibinfo{author}{\bibfnamefont{J.}~\bibnamefont{Xu}},
  \bibinfo{author}{\bibfnamefont{M.}~\bibnamefont{Kurita}},
  \bibinfo{author}{\bibfnamefont{K.}~\bibnamefont{Kato}}, \bibnamefont{and}
  \bibinfo{author}{\bibfnamefont{K.}~\bibnamefont{Adachi}},
  \bibinfo{journal}{Tribology Letters} \textbf{\bibinfo{volume}{15}},
  \bibinfo{pages}{91} (\bibinfo{year}{2003}).

\bibitem[{\citenamefont{Gao et~al.}(1998)\citenamefont{Gao, Luedtke, and
  Landman}}]{gao98}
\bibinfo{author}{\bibfnamefont{J.}~\bibnamefont{Gao}},
  \bibinfo{author}{\bibfnamefont{W.}~\bibnamefont{Luedtke}}, \bibnamefont{and}
  \bibinfo{author}{\bibfnamefont{U.}~\bibnamefont{Landman}},
  \bibinfo{journal}{J. Phys. Chem. B} \textbf{\bibinfo{volume}{102}},
  \bibinfo{pages}{5033} (\bibinfo{year}{1998}).

\bibitem[{\citenamefont{Rozman et~al.}(1998)\citenamefont{Rozman, Urbakh, and
  Klafter}}]{rozman98}
\bibinfo{author}{\bibfnamefont{M.~G.} \bibnamefont{Rozman}},
  \bibinfo{author}{\bibfnamefont{M.}~\bibnamefont{Urbakh}}, \bibnamefont{and}
  \bibinfo{author}{\bibfnamefont{J.}~\bibnamefont{Klafter}},
  \bibinfo{journal}{Phys. Rev. E} \textbf{\bibinfo{volume}{57}},
  \bibinfo{pages}{7340} (\bibinfo{year}{1998}).

\bibitem[{\citenamefont{Zaloj et~al.}(1999)\citenamefont{Zaloj, Urbakh, and
  Klafter}}]{zaloj99}
\bibinfo{author}{\bibfnamefont{V.}~\bibnamefont{Zaloj}},
  \bibinfo{author}{\bibfnamefont{M.}~\bibnamefont{Urbakh}}, \bibnamefont{and}
  \bibinfo{author}{\bibfnamefont{J.}~\bibnamefont{Klafter}},
  \bibinfo{journal}{Phys. Rev. Lett.} \textbf{\bibinfo{volume}{82}},
  \bibinfo{pages}{4823} (\bibinfo{year}{1999}).

\bibitem[{\citenamefont{Tshiprut et~al.}(2005)\citenamefont{Tshiprut, Filippov,
  and Urbakh}}]{tshiprut05}
\bibinfo{author}{\bibfnamefont{Z.}~\bibnamefont{Tshiprut}},
  \bibinfo{author}{\bibfnamefont{A.~E.} \bibnamefont{Filippov}},
  \bibnamefont{and} \bibinfo{author}{\bibfnamefont{M.}~\bibnamefont{Urbakh}},
  \bibinfo{journal}{Phys. Rev. Lett.} \textbf{\bibinfo{volume}{95}},
  \bibinfo{pages}{016101} (\bibinfo{year}{2005}).

\bibitem[{\citenamefont{Johnson and Jia}(2005)}]{johnson05}
\bibinfo{author}{\bibfnamefont{P.~A.} \bibnamefont{Johnson}} \bibnamefont{and}
  \bibinfo{author}{\bibfnamefont{X.}~\bibnamefont{Jia}},
  \bibinfo{journal}{Nature} \textbf{\bibinfo{volume}{437}},
  \bibinfo{pages}{871} (\bibinfo{year}{2005}).

\bibitem[{\citenamefont{Johnson et~al.}(2008)\citenamefont{Johnson, Savage,
  Knuth, Gomberg, and Marone}}]{johnson08}
\bibinfo{author}{\bibfnamefont{P.~A.} \bibnamefont{Johnson}},
  \bibinfo{author}{\bibfnamefont{H.}~\bibnamefont{Savage}},
  \bibinfo{author}{\bibfnamefont{M.}~\bibnamefont{Knuth}},
  \bibinfo{author}{\bibfnamefont{J.}~\bibnamefont{Gomberg}}, \bibnamefont{and}
  \bibinfo{author}{\bibfnamefont{C.}~\bibnamefont{Marone}},
  \bibinfo{journal}{Nature} \textbf{\bibinfo{volume}{451}}, \bibinfo{pages}{57}
  (\bibinfo{year}{2008}).

\bibitem[{\citenamefont{Stacey et~al.}(2005)\citenamefont{Stacey, Gomberg, and
  Cocco}}]{stacey05}
\bibinfo{author}{\bibfnamefont{S.}~\bibnamefont{Stacey}},
  \bibinfo{author}{\bibfnamefont{J.}~\bibnamefont{Gomberg}}, \bibnamefont{and}
  \bibinfo{author}{\bibfnamefont{M.}~\bibnamefont{Cocco}}, \bibinfo{journal}{J.
  Geophys. Res.} \textbf{\bibinfo{volume}{110}}, \bibinfo{pages}{B05501}
  (\bibinfo{year}{2005}).

\bibitem[{\citenamefont{Braun and Peyrard}(2001)}]{braun01}
\bibinfo{author}{\bibfnamefont{O.~M.} \bibnamefont{Braun}} \bibnamefont{and}
  \bibinfo{author}{\bibfnamefont{M.}~\bibnamefont{Peyrard}},
  \bibinfo{journal}{Phys. Rev. E} \textbf{\bibinfo{volume}{63}},
  \bibinfo{pages}{046110} (\bibinfo{year}{2001}).

\bibitem[{\citenamefont{Braun and Kivshar}(2004)}]{braun04}
\bibinfo{author}{\bibfnamefont{O.~M.} \bibnamefont{Braun}} \bibnamefont{and}
  \bibinfo{author}{\bibfnamefont{Y.~S.} \bibnamefont{Kivshar}},
  \emph{\bibinfo{title}{The Frenkel-Kontorova Model: Concepts, Methods, and
  Applications}} (\bibinfo{publisher}{Springer-Verlag, Berlin},
  \bibinfo{year}{2004}).

\end{thebibliography}

\end{document}